\documentclass[preprintnumbers,superscriptaddress,amsmath,amssymb]{revtex4}
\usepackage{graphicx}
\def\L{{\mathcal L}}

\def\L{\mathcal L}

\def\a{\alpha}
\def\b{\beta}
\def\l{\lambda}

\begin{document}

\title{Charged Higgs production at photon colliders in 2HDM-III}
\author{Roberto Mart\'\i nez}
\author{J-Alexis Rodr\'\i guez}
\email{jarodriguezl@unal.edu.co}
\author{Susana S\'anchez}
\email{smsanchezn@unal.edu.co}
\

\affiliation{Department of Physics, Universidad Nacional de Colombia}
\altaffiliation{Ciudad Universitaria, Bogot\'a, Colombia}


\begin{abstract}
We study charged Higgs production in the process $\gamma\gamma\to A^0\to W^- H^+$. The processes $\gamma\gamma\to A^0$ are loop mediated in a 2HDM. This is due to the fact that photons only couple directly to charged particles and the Higgs only couples to particles with mass acquired via Higgs mechanism. Although in MSSM the contribution  from the process  $\gamma\gamma\to A^0$ is too small, it has been found that in a more general 2HDM it could be enhanced. On the other hand, the boson $A^0$ can decay in $W^- H^+$ at tree level and the charged Higgs can decay in fermions. So, the whole process under study is $\gamma\gamma\to A^0\to (W^-\to l\nu) (H^+\to f_if_j)$ in 2HDM-III. Evidence about charged Higgs existence could demonstrate that structure of the Higgs sector has several multiplets.
\end{abstract}


\maketitle

\section{Introduction}

The Standard Model of particle physics (SM) has been succesful in explaining experimental data so far. However, the Higgs sector of the model is experimentally unknown. This fact has lead many theorists to suggest that the sector could be non-minimal, and today it is common to study extensions of the SM scalar sector. The simplest extension is the Two Higgs Doublet Model (2HDM)\cite{mssm,HHG}, which involves two scalar doublets in the process of electroweak symmetry breaking. After the symmetry breaking, there are five physical Higgs particles: two charged Higgs $H^\pm$, two CP-even $H^0$, $h^0$, and one CP-odd $A^0$. The charged particles $H^\pm$ are characteristic of the models with two Higgs doublets and its discovery would be a clear signal of physics beyond the SM.\\

A promising alternative for the search of new physics lies on the collisions of photons that will be present at ILC (International Linear Collider). The ILC  is the next greatest project to be developed after the LHC (Large Hadron Collider). The photons will be produced by Compton retrodispersion, and earlier studies show that production of neutral scalars by photon collisions present a considerable probability of detection. Besides, photons couple directly to charged particles, so $\gamma\gamma$ high energy collisions could provide a better understanding of several aspects of the SM and its extensions\cite{rdr}.\\



In this work we study the process $\gamma\gamma\to A^0\to (W^-\to l\nu) (H^+\to f_if_j)$ in the frame of a general 2HDM. The first section presents a brief overview of the 2HDM, focusing in the third type (2HDM-III). The next section contains the expressions used in the calculations. The third section contains the obtained results and finally the conclusions are drawn.

\section{The Two Higgs Doublet Model type III}

The minimal extension of the Higgs sector of the SM consists in adding a second scalar doublet with the same quantum numbers than the  first one\cite{HHG,RDiaz}. We denote them by:

\begin{eqnarray}
\Phi_1=
\left(\begin{array}{cc}
\phi_1^+\\
\phi_1^0
\end{array}\right)
\hspace*{1cm}
\Phi_2=
\left(\begin{array}{cc}
\phi_2^+\\
\phi_2^0
\end{array}\right),
\end{eqnarray}

with hypercharge $Y_{\Phi_1}=Y_{\Phi_2}=1$. Both doublets acquire vacuum expectation values different from zero:

\begin{equation}
 \left<\Phi_1\right>_0=\frac{v_1}{\sqrt{2}},\hspace*{1cm}
 \left<\Phi_2\right>_0=\frac{v_2}{\sqrt{2}}.
\end{equation}

The most general gauge invariant lagrangian which couples both Higgs fields with fermions is:

\begin{eqnarray}
-\L_Y&=&
\eta_{ij}^{U,0}\bar Q_{iL}^0\tilde\Phi_1U_{jR}^0
+\eta_{ij}^{D,0}\bar Q_{iL}^0\Phi_1D_{jR}^0\nonumber\\
&+&\xi_{ij}^{U,0}\bar Q_{iL}^0\tilde\Phi_2U_{jR}^0
+\xi_{ij}^{D,0}\bar Q_{iL}^0\Phi_2D_{jR}^0\nonumber\\
&+&\eta_{ij}^{E,0}\bar l_{iL}^0\tilde\Phi_1E_{jR}^0
+\xi_{ij}^{E,0}\bar l_{iL}^0\Phi_2E_{jR}^0+h.c.
\end{eqnarray}

where $\eta^{U,D}$ and $\xi^{U,D}$ are non-diagonal mixing matrices $3\times3$, $\tilde\Phi_i=i\sigma_2\Phi_i$, $(U,D)_R$ are right-handed fermion singlets, $Q_L$ are left-handed fermion doublets, and the index 0 indicates that the fields are not mass eigenstates.\\

In the most general case, both Higgs doublets couple with the up and down sectors, and therefore they contribute simultaneously in the proccess of mass generation for quarks. This case leads to FCNC (Flavor Changing Neutral Currents) at tree level, because it is impossible to diagonalize simultaneously both matrices $\eta$ and $\xi$. This general case is known as 2HDM type III. However, FCNC proccesses at tree level are highly supressed by the experiment. In order to avoid their existence, Glashow and Weinberg\cite{RDiaz} designed the following set of discrete symmetries:

\begin{eqnarray}
&\Phi_1\to\Phi_1 \mbox{ and } \Phi_2\to-\Phi_2,\nonumber\\
&D_{jR}\to\mp D_{jR} \mbox{ and } U_{jR}\to - U_{jR}.
\end{eqnarray}

The condition of invariance under this discrete symmetry leads to two cases:

\begin{itemize}
\item By using $D_{jR}\to -D_{jR}$, matrices $\eta^{U,0}$ and $\eta^{D,0}$ have to be eliminated from the lagrangian. In this case $\Phi_1$ decouples in the Yukawa sector and only $\Phi_2$ gives masses to sectors up and down. This case is known as 2HDM type I.
\item By using $D_{jR}\to D_{jR}$, matrices $\eta^{U,0}$ and $\xi^{D,0}$ must be eliminated from the lagrangian. In this case $\Phi_1$ couples to the down sector and $\Phi_2$ gives masses to up sector. This case is known as 2HDM type II.
\end{itemize}

The 2HDM-III is the only 2HDM that allows FCNC proccesses at tree level. Precission tests of the SM model show a great aggreement with the FCNC parameters, except for the phenomenon of neutrino oscillation. Besides, the FCNC processes don't seem to violate any fundamental law. We study the 2HDM-III because it has a more rich phenomenology, and it is possible to find the first two types as limit cases of this one.\\

In the 2HDM-III, a rotation of the scalar fields does not change the physical content of the model\cite{RDiaz}. This rotation can get rid of the VEV of one doublet. If we take $\left<\Phi_2\right>=0$, it is found that $\tan\b=0$. This is known as the {\bf fundamental parameterization}. We denote the VEV of the first doublet as $\left<\Phi_1\right>=v$.\\

For a better study of the FCNC processes, Cheng, Sher and Yuang (CSY) propose an anzats for the Yukawa matrices such that

\begin{equation}
 \xi_{ij} = \frac{\sqrt{m_im_j}}{v}\l_{ij}
\end{equation}

This anzats obeys to the fact that couplings between fermions and the Higgs particle in the SM are proportional to the mass of the fermion. Parameters $\l_{ij}$ could change the hierarchy between fermionic couplings and because of this it is expected that they would be $\sim 1$. Restrictions over parameters have been obtained in references \cite{RDiaz,bounds,jimenez}. The most relevant are:

\begin{eqnarray}
\xi_{\mu\tau}^2 &\in&[7.62 \times 10^{-4} ; 4.44 \times 10^{-2}]\nonumber\\
\xi_{\tau\tau} &\in&[-1.8 \times 10^{-2} ; 2.2 \times 10^{-2}]\nonumber\\
\xi_{\mu\mu} &\in& [-0.12;0.12]\nonumber\\
\xi_{\mu e} &\in& [-0.39;0.39]\nonumber\\
\l_{bb} &\in& [-6;6]\nonumber\\
\l_{tt} &\in& [-\sqrt{8};\sqrt{8}].\label{restrictions}
\end{eqnarray}

\section{The process $\gamma\gamma\to A^0 \to W^+H^-\to l\nu f_if_j$}

\begin{figure}[htb]
\centering
\includegraphics[width=.22\linewidth]{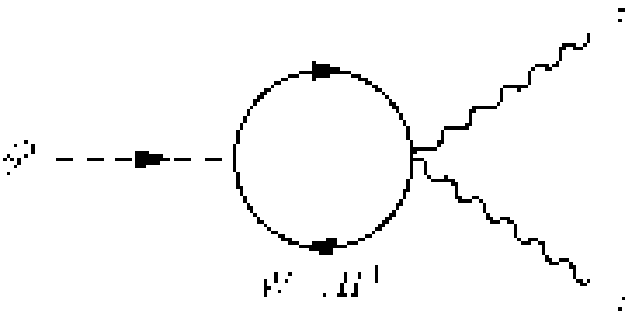}\hspace*{2cm}
\includegraphics[width=.22\linewidth]{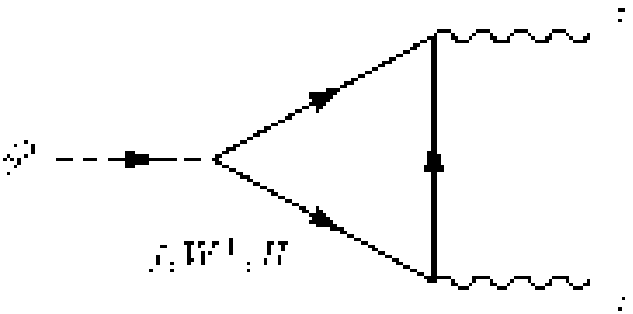}\\
\caption{Contributing diagrams in the process $\gamma\gamma\to A^0$}
\label{fig:loops}
\end{figure}

Loops contributing in neutral Higgs production are shown in Figure \ref{fig:loops} and the decay $A^0\to H^\pm W^\mp$ exists at tree level in the framework of the 2HDM-III. The process $H^-\to q_i \bar{q_j}$ has been studied in 2HDM-III\cite{hcardenas}, and under the restrictions mentioned above, it has been found that the most relevant decay is $H^-\to t\bar b$.\\

The decay width of the process $\gamma\gamma\to A^0$ is given by

\begin{eqnarray}
\Gamma\left(\gamma\gamma\to A^0\right) = \frac{\alpha^2 g^2}{1024\pi^3}\frac{m_{A^0}^3}{m_W^2}\left| \sum_i N_Ce_i^2F(\tau)R_i^{A^0} \right|^2.
\end{eqnarray}

The factor $R_i^{A^0}$ is the relative coupling between the 2HDM-III and the SM, the kinematical factor is $\tau=4m_i^2/m_{A^0}^2$, the function $F(\tau)$ is defined as:

\begin{equation}
F(\tau)=-2\tau\left(1+(1-\tau)f(\tau)\right),
\end{equation}

and the function $f$ is:

\begin{eqnarray}
f(\tau)=\left\lbrace
\begin{array}{ll}
-\frac14\left|\mbox{Ln}\left(\frac{1+\sqrt{1-\tau}}{1-\sqrt{1-\tau}}\right)-i\pi\right|^2&\tau<1\\
\mbox{ArcSin}\left(\sqrt{\frac{1}{\tau}}\right)^2&\tau\ge 1
\end{array}
\right..
\end{eqnarray}

\begin{figure}[htb]
\centering
\includegraphics[width=.5\linewidth]{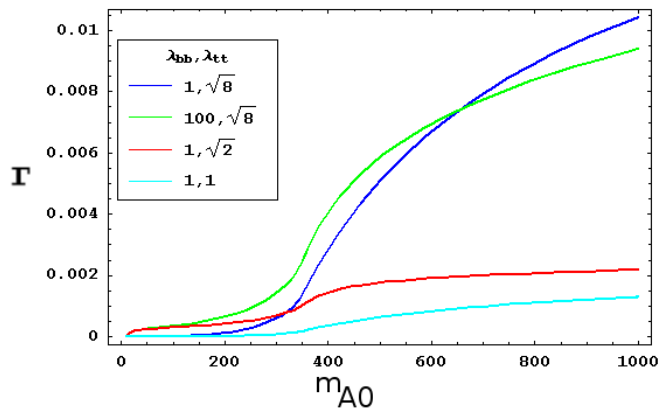}
\caption{\small Decay width of the process $\gamma\gamma\to A^0$ to one loop for $\l_{bb}=1$ and $\l_{tt}=\sqrt{8}$}
\label{fig:decaywidth}
\end{figure}

Figure \ref{fig:decaywidth} shows the decay width of the process $\gamma\gamma\to A^0$ for several values of the parameters $\l_{bb}$ and $\l_{tt}$ according to the restrictions mentioned in equation \ref{restrictions}. It is found that changes in $\l_{bb}$ do not have a big impact on the results, while changes in $\l_{tt}$ are considerable. It is found also that the decay width increases with the $A^0$ mass. So we will consider $\l_{bb}=6$ and values for $m_{A^0}>600$GeV.\\

The decay width for the process $A^0\to W^-H^+$ is given by

\begin{eqnarray}
\Gamma(A^0\to W^-H^+)=\frac{\cos^2\a m_{H^+}^3}{64\pi v^2}\times\hspace*{2.5cm}\nonumber\\
\left[\sqrt{1-\left(\frac{m_{W^-}+m_{A^0}}{m_{H^+}}\right)^2}\right.
\left.\sqrt{1-\left(\frac{m_{W^-}-m_{A^0}}{m_{H^+}}\right)^2}\right]^3,
\end{eqnarray}

where $\a$ is the mixing angle of the Higgs eigenstates, $v$ is the vacuum expectation value as defined in the fundamental parameterization.\\

It has been found that the most relevant decay of the charged Higgs into fermions for this model is the decay into $t\bar b$ quarks\cite{hcardenas}. This is given by

\begin{eqnarray}
\Gamma(H^+\to t\bar b) = \frac{3m_{H^+}^2K_{tb}^2}{16\pi v^2}\times\hspace*{3.5cm}\nonumber\\
\left[\left(1-\frac{m_t^2+m_b^2}{m_{H^+}^2}\right)-4\l_{tt}\l_{bb}\frac{m_t^2m_b^2}{m_{H^+}^2}\right]\times
\left|\vec p_H(m_t,m_b)\right|,
\end{eqnarray}

where $K_{tb}$ is the CKM matrix element, and $\left|\vec p_H(1,2)\right|$ is defined as:

\begin{eqnarray}
\left|\vec p_H(m_1,m_2)\right|=\hspace*{5cm}\nonumber\\
\sqrt{1-\left(\frac{m_{1}+m_{2}}{m_{H}}\right)^2}
\sqrt{1-\left(\frac{m_{1}-m_{2}}{m_{H}}\right)^2}.
\end{eqnarray}

The cross section for the whole process is calculated as

\begin{eqnarray}
\sigma(\gamma\gamma\to A^0\to \tau\nu_{\tau}t\bar b)=\hspace*{4cm}\\
8\pi\frac{\Gamma(\gamma\gamma\to A^0)\Gamma(A^0\to\tau\nu_{\tau}t\bar b)}{(E_{\gamma\gamma}^2-m_{A^0}^2)^2+\Gamma_{A^0}^2m_{A^0}^2}g(\l\l')\nonumber
\end{eqnarray}

Figures \ref{fig:sigma600} and \ref{fig:sigma800} show the cross section for the process $\gamma\gamma \to A^0\to W^+H^- \to (\tau\nu_\tau)(t\bar b)$, using $E_{\gamma\gamma}=1000$GeV.

\section{Conclussions}

\begin{figure}[htb]
\centering
\includegraphics[width=.5\linewidth]{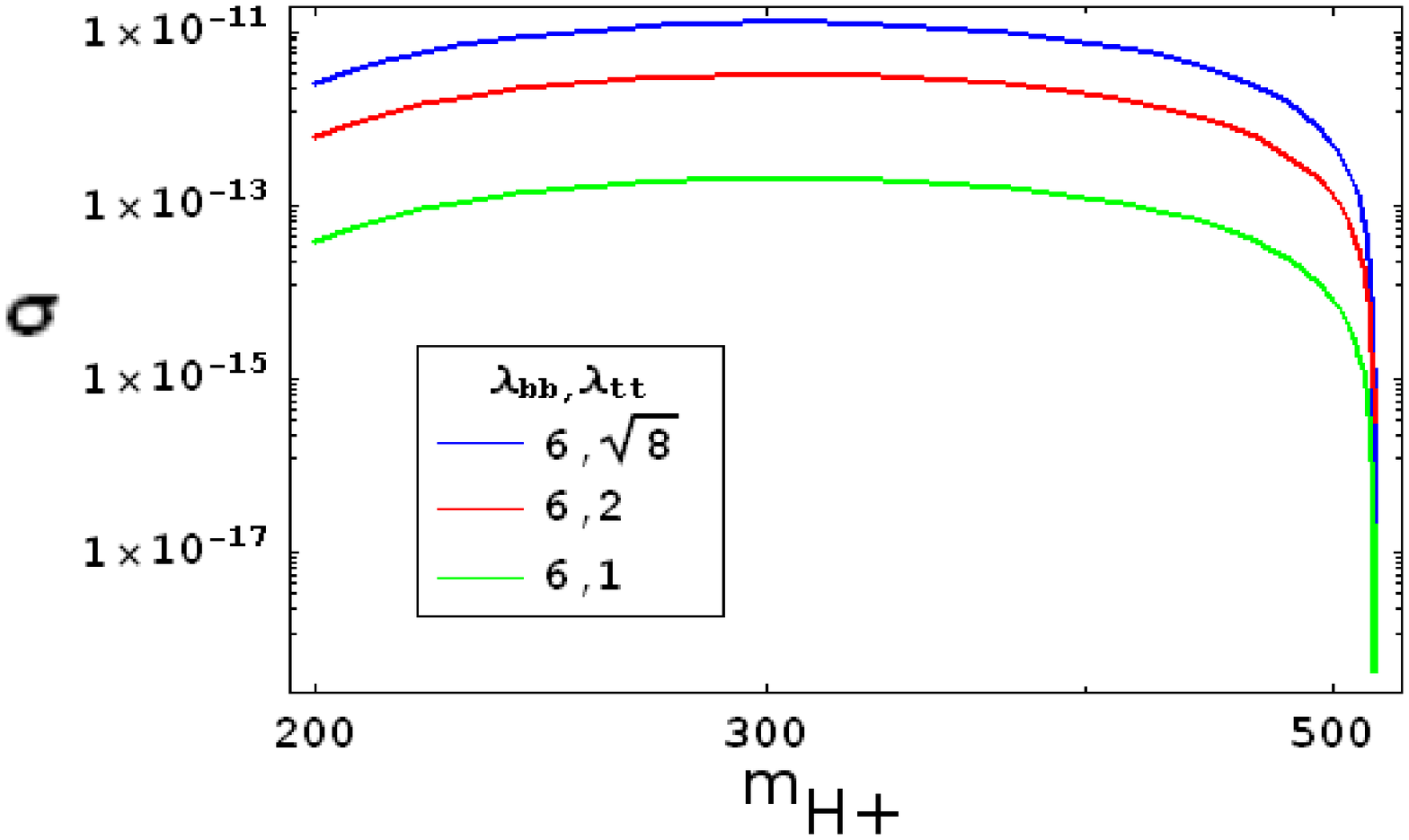}
\caption{\small Cross section (in barns) of the process $\gamma\gamma\to A^0\to W^-H^+ \to e\nu_et\bar b$ for different values of $\l_{tt}$, using $m_{A^0}=600$GeV and $\l_{bb}=6$}
\label{fig:sigma600}
\end{figure}

\begin{figure}[htb]
\centering
\includegraphics[width=.5\linewidth]{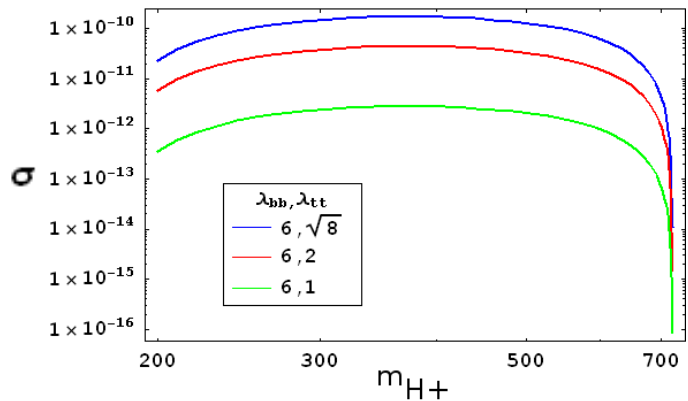}
\caption{\small Cross section (in barns) of the process $\gamma\gamma\to A^0\to W^-H^+ \to e\nu_et\bar b$ for different values of $\l_{tt}$, using $m_{A^0}=800$GeV and $\l_{bb}=6$}
\label{fig:sigma800}
\end{figure}

We found the cross section for the process $\gamma\gamma \to A^0\to W^+H^- \to (\tau\nu_\tau)(t\bar b)$ in the frame of the 2HDM-III. Results are shown in Figures \ref{fig:sigma600} and \ref{fig:sigma800}. It is found that the cross section takes values between 1pb and 10pb for the parameters $\l_{bb}=6$, $\l_{tt}=\sqrt{8}$ and $m_{A^0}=600$GeV. For a higher value of $m_{A^0}$ we get higher values of the cross section, between 10pb and 100pb. For lower values of the parameter $\l_{tt}$, we still get cross section values between 0.1pb and 10pb.\\

Finally, we can say that for two Higgs doublets models type III, the contribution of the process $\gamma\gamma\to A^0\to W^-H^+ \to e\nu_et\bar b$ is important, even though it is loop mediated. Earlier studies show that the contribution of this kind of processes in models such as MSSM is null\cite{asakawa}. The presence of this process would help to diferentiate between the MSSM and a more general 2HDM. Besides, evidence of the charged Higgs existence would demonstrate the multiple doublets structure of the Higgs sector.\\

\begin{acknowledgments}
R.M acknowledges to Banco de la Rep\'ublica for the financial support in the development of this work.
\end{acknowledgments}

\end{document}